# Optimization with Demand Oracles

Ashwinkumar Badanidiyuru
Department of Computer Science
Cornell Unversity
ashwin85@cs.cornell.edu

Shahar Dobzinski
Department of Computer Science
Cornell Unversity
shahar@cs.cornell.edu

Sigal Oren
Department of Computer Science
Cornell Unversity
sigal@cs.cornell.edu

November 7, 2018

## Abstract

We study *combinatorial procurement auctions*, where a buyer with a valuation function $v$ and budget $B$ wishes to buy a set of items. Each item $i$ has a cost $c_i$ and the buyer is interested in a set $S$ that maximizes $v(S)$ subject to $\Sigma_{i \in S} c_i \leq B$. Special cases of combinatorial procurement auctions are classical problems from submodular optimization. In particular, when the costs are all equal (*cardinality constraint*), a classic result by Nemhauser et al shows that the greedy algorithm provides an $\frac{e}{e-1}$ approximation.

Motivated by many papers that utilize demand queries to elicit the preferences of agents in economic settings, we develop algorithms that guarantee improved approximation ratios in the presence of demand oracles. We are able to break the $\frac{e}{e-1}$ barrier: we present algorithms that use only polynomially many demand queries and have approximation ratios of $\frac{9}{8} + \epsilon$ for the general problem and $\frac{9}{8}$ for maximization subject to a cardinality constraint.

We also consider the more general class of subadditive valuations. We present algorithms that obtain an approximation ratio of $2 + \epsilon$ for the general problem and $2$ for maximization subject to a cardinality constraint. We guarantee these approximation ratios even when the valuations are non-monotone. We show that these ratios are essentially optimal, in the sense that for any constant $\epsilon > 0$, obtaining an approximation ratio of $2 - \epsilon$ requires exponentially many demand queries.

# 1   Introduction

We study the following *combinatorial procurement auction* problem: a buyer with a valuation $v$ and budget $B$ wishes to purchase a set of items $S$, where each item $i$ has a cost $c_i$. The buyer is interested in maximizing his value $v(S)$ while not overspending ($\Sigma_{i \in S} c_i \leq B$).

*Truthful* mechanisms for various variations of this problem were studied in [26] and its followup [9]. In this paper we study the problem from a pure combinatorial optimization point of view. This is analogous to the combinatorial auctions literature, where one branch studies incentives issues (e.g., [8, 20, 6, 2]), and the other studies the problem as a pure combinatorial optimization problem ignoring incentives (e.g., [22, 16, 27, 7, 11, 13, 10]).

We focus on cases where the valuation function $v$ is submodular (for each two sets $S$ and $T$, $v(S)+v(T) \geq v(S \cup T)+v(S \cap T)$) or subadditive (for each two sets $S$ and $T$, $v(S)+v(T) \geq v(S \cup T)$). Submodular valuations capture cases where the buyer exhibits decreasing marginal utilities, and subadditive valuations capture complement freeness. See, e.g., [22], for a more elaborate discussion.

When $v$ is submodular a combinatorial procurement auction is a reformulation of a classical problem from submodular optimization, maximization subject to a knapsack constraint: each item $i$ has a cost $c_i$ and the goal is to find a maximum-value set $S$ such that $\Sigma_{i \in S} c_i \leq B$, for a given budget $B$. A special case is maximization subject to a cardinality constraint: find a set $S$ of size $k$ with the highest value. Many other generalizations and variants of this problem were studied (e.g., [14, 27, 17, 21, 18, 5]).

We want our algorithms to run in time polynomial in $n$, the number of items. However, the valuation function is an object of size $2^m$. The combinatorial auctions literature therefore usually assumes that access to the valuation is done via an oracle. Two types of queries were extensively studied: *value queries* (given a set $S$, return $v(S)$) and *demand queries* (given prices $p_1, \ldots, p_m$ return a set $S$ such that $S \in \arg\max_T v(T) - \Sigma_{j \in T} p_j$). Here we use demand queries to solve the combinatorial procurement auctions problem. This path was already taken in [9], but let us mention two reasons:

- **Economic interpretation:** many algorithms for economic settings, either truthful or not truthful, assume that the valuations are accessed via demand oracles (e.g., [13, 11, 7, 8, 1, 20]). As was argued extensively in the literature (e.g., [4, 3, 25, 19, 24]) demand queries are a natural way for agents to express their preferences.

- **Bypassing impossibility results:** it is known that polynomially many *value* queries cannot guarantee an approximation ratio of $n^{\frac{1}{2}-\epsilon}$ even for optimization subject to a cardinality constraint if the valuation of the buyer is subadditive [26, 7][1]. Hence we *must* use stronger oracles to achieve reasonable approximation ratios.

The classical result in submodular optimization shows that the greedy algorithm provides an approximation ratio of $\frac{e}{e-1}$ [14] for optimization subject to a cardinality constraint, and that this is the best possible with a polynomial number of value queries [23, 12]. We show that demand queries allow us to break the $\frac{e}{e-1}$ barrier:

**Theorem:** There exists a $\frac{9}{8}$-approximation algorithm for the problem of maximizing a monotone submodular function subject to a cardinality constraint that makes a polynomial number of demand queries. For the problem of maximizing a monotone submodular function subject to a knapsack constraint, for every constant $\epsilon > 0$, there exists a $(\frac{9}{8} + \epsilon)$-approximation algorithm that makes a polynomial number of demand queries.

---

[1]In fact this bound holds even if the valuation is fractionally subadditive – a valuation is fractionally subadditive (a.k.a. XOS) if it is the maximum of several additive valuations. A formal definition will be presented later.

We start by presenting a natural linear program. The dual of the linear program has exponential many constraints but only polynomially many variables, so we show that the dual can be solved via the ellipsoid method, using a demand query as a separation oracle[2]. We show various structural properties of optimal solutions to the LP, e.g., there are at most two sets in their support. We now face an additional challenge: in general, none of these sets (or a part of them, in case they violate the cardinality constraint) provides a good enough approximation. The key step is showing that by taking one set and augmenting it using the other set we get a new combined set that does provide the specified approximation ratio.

Next, we consider the class of subadditive valuations. In [9] a $(2 + \epsilon)$ approximation was obtained for maximization subject to a cardinality constraint. On top of slightly improving this ratio, we present the first constant-approximation algorithm for maximization subject to a knapsack constraint, improving over the $O(\log m)$ ratio in [9]:

**Theorem:** There exists a 2-approximation algorithm for the problem of maximizing a subadditive function subject to a cardinality constraint that makes a polynomial number of demand queries. For the problem of maximizing a subadditive function subject to a knapsack constraint, for every constant $\epsilon > 0$, there exists a $(2 + \epsilon)$-approximation algorithm that makes a polynomial number of demand queries.

These algorithms do not assume that the valuation is monotone. In particular, the algorithms guarantee the specified approximation ratio also for non-monotone submodular valuations (a class that received much attention recently – see, e.g., [21, 18, 15, 5]). In fact, we once again break the lower bound for approximation with value queries: it is known that a 2.03-approximation algorithm for maximizing non-monotone submodular function must use exponentially many value queries [15].

We also show how to obtain purely combinatorial algorithms via a certain type of a natural "ascending auction", possibly with an additional augmentation step. An exciting direction is to analyze these auctions and similar ones from a more economic point of view, although we do not push this direction further in this paper.

Our bounds have another implication: they enable us to construct a "monotone estimator" (in the language of [9]) and improve the best known approximation ratio of a deterministic *truthful* mechanism for maximizing subadditive valuation subject to a knapsack constraint to $O(\log^2 m)$. The best previously known bound for this setting was $O(\log^3 m)$ [9].

Are our bounds optimal? We provide an example with a matching integrality gap of $\frac{9}{8}$ for optimizing monotone submodular valuations. For non-monotone submodular valuations we present an example with a matching integrality gap of 2. Moreover, we show that for subadditive valuations our results are optimal (in fact, they are optimal even if the function is known to be fractionally subadditive):

**Theorem:** Fix some constant $\epsilon > 0$. Let $A$ be a (possibly randomized) $(2 - \epsilon)$-approximation algorithm for maximizing a fractionally subadditive function subject to a cardinality constraint. Then, $A$ makes exponentially many demand queries.

We note that proving limits on the power of demand queries requires developing significant amount of novel machinery[3]. The idea is to start with some specific valuation $v$, and obtain a valuation $v_T$ by "planting" some bundle $T$ of size $k$ with high value. We would like to show that determining whether such planting occurred requires an exponential number of demand queries. The challenge here is that a single demand query can verify whether the valuation is $v_T$ for many bundles $T$ simultaneously. Nevertheless, we show that the power of a single demand query is limited: for every demand query there exists a set of items (relatively small but of significant size) that is contained in all bundles $T$ that the demand query simultaneously verifies. This enables us to upper bound the number of bundles that can be simultaneously verified by a single demand query, and suffices to derive the theorem. We also use this technique to rule out an FPTAS for maximizing a submodular function subject to a knapsack constraint.

---

[2]This first step is very similar to the use of demand oracles in algorithms for combinatorial auctions, where demand oracles are used to solve the LP [7, 11, 13], but the similarity between the algorithms ends here.

[3]The most relevant lower bound that applies specifically to demand queries is that of Nisan and Segal [24].

**Open Questions**

In this paper we initiated the systematic study of optimization with demand queries. Let us now mention several intriguing questions that we leave open. The first obvious one is to determine the possible approximation ratio when the valuation of the buyer is submodular, both in the monotone and in the non-monotone cases. It is also interesting to understand the possible approximation ratios using different constraints. Examples include optimizing submodular functions subject to matroid constraint [14, 27], multiple knapsacks [17, 18] and their combination [21, 5]. What is the approximation ratio possible if demand queries are available in all these settings? We make a first step in this direction by providing an $O(k)$ approximation algorithm for optimizing subadditive function subject to $k$ knapsack constraints. We do not know neither whether this is the optimal ratio possible nor how to construct good algorithms for the other settings.

**Paper Organization**

Section 3 presents the LP for the problem and proves some structural properties of optimal solutions. In Section 4 we give our $\frac{9}{8}$-approximation algorithms for submodular valuations. Section 5 provides approximation algorithms for subadditive valuations. We show that the algorithms for subadditive valuations are in fact optimal in Section 6. The proof that there is no FPTAS for submodular valuations subject to a knapsack constraint is also in that section.

## 2 Preliminaries

**Valuations and Problems Definition**

Let $M$ be a set of items, $|M| = m$. Let $v : 2^M \to \mathbb{R}$ be a set function. We assume that $v(\emptyset) = 0$. $v$ is *monotone non-decreasing* if for each $S \subseteq T$, $v(S) \leq v(T)$.

In this paper we are mainly interested in the following two problems[4]: in the *maximization subject to cardinality constraint* problem we are given a number $k$, and we want to find the largest value set $S$, $|S| = k$. In the *maximization subject to budget constraint problem* we are given a budget $B$ and a cost $c_i$ for each item $i$. The goal is to find the largest-value set $S$, such that $\Sigma_{j \in S} c_j \leq B$. We sometimes use the name *maximization subject to a knapsack constraint* for this problem.

We consider this problem under various restrictions on the valuations. We say that $v$ is *submodular* if for every $S, T \subseteq M$ we have that $v(S) + v(T) \geq v(S \cup T) + v(S \cap T)$. $v$ is *subadditive* if for every $S, T \subseteq M$ we have that $v(S) + v(T) \geq v(S \cup T)$. A valuation is *additive* if $v(S) = \Sigma_{j \in S} v(\{j\})$. Notice that every submodular valuation is also subadditive, and every additive valuation is submodular.

---

However, the setting of [24] is much simpler, and their ideas do not seem to be applicable in our case.

[4]We also consider some generalizations, which will be defined in the proper subsections.

Another type of valuations that we consider is fractionally subadditive (or XOS). A valuation is *XOS* if there exist additive valuations $v_1, \ldots, v_l$ such that $v(S) = \max_i v_i(S)$ (we say that $i$ is a *maximizing clause* of $S$). It is known [22] that every XOS valuation is subadditive, and that every monotone submodular valuation is XOS, and that there are subadditive valuations that are not XOS, and XOS valuations that are not submodular.

We sometimes use the notation $v(S|T)$ to denote $v(S \cup T) - v(T)$. For the problem of maximizing a function subject to budget constraint, we sometimes use the notation $C(S) = \Sigma_{j \in S} c_j$.

**Demand Queries**

Given prices $p_1, \ldots, p_m$, a *demand query* returns a bundle $S$, $S \in \arg\max_T v(T) - \Sigma_{j \in T} p_j$. The set of bundles that achieve the max is called the *demand set* of the query. Our algorithms provide the guaranteed approximation ratio without making any assumption about which specific bundle is returned from the demand set. For lower bounds, as pointed out in [24], some unnatural tie-breaking rules may supply unrealistic information about the valuation. Therefore, as in [24], our lower bound assumes any tie-breaking rule that does not depend on the valuation, e.g., return the lexicographically-first bundle in the demand set.

Another type of well-studied query is *value query*: given a set $S$, return $v(S)$. It is known [4] that a value query can be simulated by polynomially many demand queries (but exponentially many value queries may be required to simulate a single demand query). This paper concentrates in designing algorithm with polynomially many demand queries, so we freely assume that we have access to value queries.

## 3 The Structure of the LP

Our algorithms are based on finding good roundings of fractional solutions. These fractional solutions are obtained by solving a natural linear relaxation of our problems. The key ingredient is analyzing the structural properties of optimal fractional solutions: we give limits on the number of elements in the support, analyze their costs, and give precise formulas for the weights of the elements in the support.

The LP can be solved using demand oracles. Moreover, we show that there is a combinatorial method of obtaining solutions to the LP, using a process that can be viewed as an ascending auction.

The LP is presented for the more general case of maximization subject to a budget constraint (maximization subject to a cardinality constraint is a special case where for each item $i$, $c_i = 1$ and $B = k$).

*Maximize:* $\sum_S x_S \cdot v(S)$
*Subject to:*

- $\sum_S x_S \cdot C(S) \leq B$.
- $\sum_S x_S \leq 1$.
- For each bundle $S$: $x_S \geq 0$.

Although the LP has an exponential number of variables, we show that it can still be solved using a polynomial number of demand queries:

**Proposition 3.1** *The LP can be solved using a polynomial number of demand queries.*

**Proof:** Consider the dual of the primal LP:

*Minimize:* $y \cdot B + z$
*Subject to:*

- For each bundle $S$: $z + y \cdot C(S) \geq v(S)$.
- $y \geq 0$.
- $z \geq 0$.

We use the ellipsoid method to solve the dual LP and thus obtain a solution to the primal LP. For the ellipsoid method to work we need to implement a separation oracle that finds some set $S$ that violates the constraint $y \cdot C(S) + z \geq v(S)$. To do that, consider the demand query $dq = y \cdot (c_1, \ldots, c_m)$, and let $T$ be the bundle that $dq$ returns. If $z \geq v(T) - y \cdot C(T)$ then by the definition of demand query for every other set $S$ we have $z \geq v(T) - y \cdot C(T) \geq v(S) - y \cdot C(S)$. □

A key ingredient of our algorithms is the following structural property of the LP.

**Definition 3.2** *A fractional solution is a* strict *solution if there are two variables $x_{S_1}$ and $x_{S_2}$, such that $C(S_1) \leq B$, $C(S_2) > B$, $x_{S_1} = \alpha$, and $x_{S_2} = 1 - \alpha$, where $\alpha = \frac{C(S_2) - B}{C(S_2) - C(S_1)}$.*

**Proposition 3.3** *Either there exists an optimal integral solution to the LP or there exists an optimal strict solution. Given an optimal solution to the LP, we can transform it in polynomial time to a fractional solution with the same value that is either integral or strict.*

**Proof:** Suppose that for each set $S$ in the support of the solution we have that $C(S) \leq B$. In this case we claim that there is an integral solution, since $\max(v(S)|x_S > 0) \geq \sum_S x_S v(S)$.

Consider now the other extreme case, where for each set $S$ in the support of the solution we have that $C(S) > B$. In this case we show that there is a fractional solution with exactly one variable in the support:

$$\sum_S x_S v(S) = \sum_S x_S \sum_{j \in S} c_j \frac{v(S)}{\sum_{j \in S} c_j}$$
$$= \max(\frac{v(S)}{\sum_{j \in S} c_j}|x_S > 0) \cdot \sum_S x_S \sum_{j \in S} c_j$$
$$\leq \max(\frac{v(S)}{\sum_{j \in S} c_j}|x_S > 0) \cdot B$$

Let $S_2 = \arg\max(\frac{v(S)}{\sum_{j \in S} c_j}|x_S > 0)$. In this case we set $S_1 = \emptyset$. Observe that by setting $\alpha = \frac{C(S_2) - B}{C(S_2)}$ we get a strict fractional solution with value at least equal to the value of the optimal fractional solution.

The only case that is left to handle is the case where there are two sets with $x_{S_1}, x_{S_2} > 0$ and $C(S_1) \leq B$ and $C(S_2) > B$. We show a strict fractional solution that is optimal. By complementary slackness we have that $z + y \cdot C(S_1) = v(S_1)$ and $z + y \cdot C(S_2) = v(S_2)$. Solving for $z$ and $y$ we get $z = \frac{v(S_1)C(S_2) - v(S_2)C(S_1)}{C(S_2) - C(S_1)}$ and $y = \frac{v(S_2) - v(S_1)}{C(S_2) - C(S_1)}$. Hence the value of the optimal fractional solution is $\frac{v(S_1)(C(S_2) - B) + v(S_2)(B - C(S_1))}{C(S_2) - C(S_1)}$.

Consider now the strict fractional solution $x_{S_1} = \frac{C(S_2) - B}{C(S_2) - C(S_1)}$, $x_{S_2} = \frac{B - C(S_1)}{C(S_2) - C(S_1)}$. Its value is exactly the value of the optimal fractional solution: $\frac{C(S_2) - B}{C(S_2) - C(S_1)} \cdot v(S_1) + \frac{B - C(S_1)}{C(S_2) - C(S_1)} \cdot v(S_2)$ □

We now present a combinatorial method of obtaining strict fractional solutions to the LP. An important advantage of this method is that it uses only uniform-price queries (unlike the previous LP-based approach that requires arbitrary demand queries). We make use of the following simple property of demand queries:

**Lemma 3.4** *Let $p = (p_1, ..., p_m)$. Let $S_1$ be some bundle that maximizes the profit in prices $\lambda_1 \cdot p$. Let $S_2$ be some bundle that maximizes the profit in prices $\lambda_2 \cdot p$. Then $\lambda_1 \leq \lambda_2$ if and only if $\Sigma_{j \in S_1} p_j \geq \Sigma_{j \in S_2} p_j$.*

**Proof:** By the definition of demand query we have $v(S_2) - \lambda_1 \Sigma_{j \in S_2} p_j \leq v(S_1) - \lambda_1 \Sigma_{j \in S_1} p_j$ and $v(S_1) - \lambda_2 \Sigma_{j \in S_1} p_j \leq v(S_2) - \lambda_2 \Sigma_{j \in S_2} p_j$. Adding the two inequalities and simplifying we get $(\lambda_1 - \lambda_2)(\Sigma_{j \in S_2} p_j - \Sigma_{j \in S_1} p_j) \geq 0$. This implies the lemma. □

**Definition 3.5** *$\lambda \geq 0$ is the boundary if there exists $S_1, S_2$, $C(S_1) \leq B, C(S_2) > B$, such that $S_2$ is in the demand set in prices $\lambda \cdot (c_1, \ldots, c_m)$ and for every small enough $\delta > 0$ we have that $S_1$ is in the demand set in prices $(\lambda + \delta) \cdot (c_1, \ldots, c_m)$. We say that $S_1$ and $S_2$ are two boundary bundles.*

Let us further explain the notion of boundary bundles. For simplicity, this paragraph considers only the simpler cardinality constraint. Consider some prices $\lambda \cdot (1, \ldots, 1)$. When $\lambda = 0$, the most profitable bundle consists of all items. Lemma 3.4 implies that as the value of $\lambda$ increases, the size of bundles in the demand set decreases (eventually, when $\lambda$ is big enough, the demand set will be empty). The boundary is the specific $\lambda$ for which the demand set in $\lambda \cdot (1, \ldots, 1)$ contains only bundles that violate the cardinality constraint, but for $(\lambda + \epsilon) \cdot (1, \ldots, 1)$ the demand set contains bundles that respect the cardinality constraint.

The boundary bundles (and $\lambda$) can be found by an ascending auction in which we continously[5] increase the value of $\lambda$ and obtain a profit-maximizing bundle in the current prices. The boundary is the $\lambda$ for which the prices $\lambda \cdot (c_1, \ldots, c_m)$ are the supremum of the points where supply exceeds demand. We were surprised to find that the boundary bundles define a high-value fractional solution:

**Lemma 3.6** *Let $S_1$ and $S_2$ be two boundary bundles and let $\lambda$ be the boundary. Let $\alpha = \frac{C(S_2) - B}{C(S_2) - C(S_1)}$. Then $x_{S_1} = \alpha$ and $x_{S_2} = 1 - \alpha$ is a strict solution to the LP. Moreover, let $O$ be the optimal integral solution. Then, $x_{S_1} v(S_1) + x_{S_2} v(S_2) \geq v(O)$.*

**Proof:** To see that the solution respects the budget constraint:

$$
\begin{aligned}
\alpha \cdot C(S_1) + (1 - \alpha) \cdot C(S_2) &= \frac{C(S_2) - B}{C(S_2) - C(S_1)} \cdot C(S_1) + \frac{B - C(S_1)}{C(S_2) - C(S_1)} \cdot C(S_2) \\
&= \frac{B(C(S_2) - C(S_1))}{C(S_2) - C(S_1)} \\
&= B
\end{aligned}
$$

As for the second part, $S_1$ and $S_2$ are in the demand set of their respective prices, therefore:

$$
\begin{aligned}
v(O) - (\lambda + \delta) \cdot C(O) &\leq v(S_1) - (\lambda + \delta) \cdot C(S_1) \\
v(O) - \lambda \cdot C(O) &\leq v(S_2) - \lambda \cdot C(S_2)
\end{aligned}
$$

---
[5]Of course, to obtain a discrete process one may increment $\lambda$ in some discrete amount. This results in some loss in the approximation ratios of the algorithms that we construct (the loss depends on the size of the increments).

Multiplying the first inequality by $\alpha$, the second inequality by $(1-\alpha)$, and summing up the two we get that (using $B \geq C(O)$):

$$\begin{aligned}
v(O) - (\lambda + \delta)B + \alpha\delta \cdot B &\leq \alpha \cdot v(S_1) + (1-\alpha)v(S_2) - \alpha \cdot (\lambda + \delta) \cdot C(S_1) - (1-\alpha) \cdot \lambda \cdot C(S_2) \\
&= \alpha \cdot v(S_1) + (1-\alpha)v(S_2) - \frac{C(S_2) - B}{C(S_2) - C(S_1)} \cdot (\lambda + \delta) \cdot C(S_1) \\
&\quad - \frac{B - C(S_1)}{C(S_2) - C(S_1)} \cdot \lambda \cdot C(S_2) \\
&= \alpha \cdot v(S_1) + (1-\alpha)v(S_2) - \lambda \cdot B - \frac{C(S_2) - B}{C(S_2) - C(S_1)} \cdot \delta \cdot C(S_1)
\end{aligned}$$

Rearranging both sides we get:

$$v(O) - \delta \cdot B + \alpha\delta \cdot B \leq \alpha \cdot v(S_1) + (1-\alpha)v(S_2) - \frac{C(S_2) - B}{C(S_2) - C(S_1)} \cdot \delta \cdot C(S_1)$$

Now when we let $\delta$ go to 0 we get that $x_{S_1} v(S_1) + x_{S_2} v(S_2) \geq v(O)$, as needed. □

## 4 A $\frac{9}{8}$-Approximation for Monotone Submodular Valuations

As we have shown, any strict solution to the LP with cardinality constraint contains two bundles, a "large" bundle with more than $k$ items, and a "small" bundle with at most $k$ items. A natural approach for an approximation algorithm is to select the maximum of the following two bundles: the "small" bundle, or an high-value chunk of size $k$ from the "large" bundle. Unfortunately, there are examples that show that this approach cannot provide an approximation ratio better than 2. Hence, our strategy is subtler: we start with the small bundle and grab as much value as we can from the large bundle. We show that this combined bundle provides an approximation ratio of $\frac{9}{8}$.

We also extend this algorithm to handle maximization subject to a knapsack constraint. Here the approximation ratio we get is slightly worse: $\frac{9}{8}+\epsilon$. We enumerate over all sets of high-cost items, then use a variation of our algorithm for a cardinality constraint as an algorithm for instances with only low-cost items. We complement these results by presenting an instance with an integrality gap of $\frac{9}{8}$ (in the appendix), even for maximization subject to cardinality constraint.

**Theorem 4.1** *The following algorithms exist:*

- *A $\frac{9}{8}$-approximation algorithm for the problem of maximizing a monotone submodular function subject to a cardinality constraint that makes a polynomial number of demand queries.*

- *A $(\frac{9}{8}+O(\epsilon))$-approximation algorithm for the problem of maximizing a monotone submodular valuation subject to a knapsack constraint that makes a polynomial number of demand queries, for any fixed $\epsilon > 0$.*

We start with the algorithm for the cardinality constraint. We first present the algorithm, then comment on how it can be efficiently implemented.

**The Algorithm**

1. Obtain a strict solution to the LP: $x_{S_1} = \alpha$ and $x_{S_2} = 1 - \alpha$. Let $k_1 = |S_1|$ and $k_2 = |S_2|$.

2. Find $S' \subseteq S_2$, $|S'| = k$ such that $v(S') \geq \frac{k}{k_2} v(S_2)$.

3. Find $S'' \subseteq S_2 - S_1$, $|S''| = k - k_1$, such that $v(S''|S_1) \geq \frac{k-k_1}{|S_2-S_1|}v(S_2|S_1)$.

4. Output $\max(v(S_1 \cup S''), v(S'))$.

As for the efficient implementation of the algorithm, we have already argued in Proposition 3.3 that an optimal and strict solution to the LP can be found with a polynomial number of demand queries. Alternatively, Step 1 can also be implemented combinatorially as an ascending auction: start with a price per item of 0 and increase it gradually. When supply exceeds demand, consider the boundary bundles, and obtain a strict solution as in Lemma 3.6. Steps 2 and 3 can be implemented using the following folklore lemma (see Lemma 4.5 for a proof of a more general setting).

**Lemma 4.2** *Let $v$ be a submodular valuation, $S$ be some bundle, and let $t \leq |S|$ be some integer. Then, there exists a set $S' \subseteq S$, $|S'| = t$ such that $v(S') \geq \frac{t}{|S|}v(S)$. In addition, $S$ can be found using a polynomial number of value queries.*

Step 2 follows immediately from Lemma 4.2. Step 3 follows by observing that the marginal valuation $v(\cdot|S_1)$ is submodular too, and applying Lemma 4.2 again. We are left with proving the approximation ratio of the algorithm:

**Lemma 4.3** *Let $A$ be the bundle that the algorithm outputs. Then, $v(A) \geq \frac{8}{9}(x_{S_1}v(S_1) + x_{S_2}v(S_2))$.*

**Proof:** Recall that $v(S''|S_1) \geq \frac{k-k_1}{|S_2-S_1|}v(S_2|S_1)$. That is (also using $|S_2 - S_1| \leq k_2$):

$$v(S'' \cup S_1) - v(S_1) \geq \frac{k-k_1}{k_2}(v(S_2 \cup S_1) - v(S_1))$$

Rearranging and using $v(S_2 \cup S_1) \geq v(S_2)$ (since $v$ is monotone) we have:

$$v(S'' \cup S_1) \geq \frac{k-k_1}{k_2}v(S_2) + \frac{k_1+k_2-k}{k_2}v(S_1) \qquad (1)$$

We are finally ready to prove the approximation ratio:

$$\begin{aligned}
\alpha \cdot v(S_1) + (1-\alpha) \cdot v(S_2) &= \alpha \frac{k_2}{k_1+k_2-k}\left(\frac{k_1+k_2-k}{k_2}v(S_1) + \frac{k-k_1}{k_2}v(S_2)\right) \\
&\quad + (1-\alpha - \alpha\frac{k-k_1}{k_1+k_2-k})v(S_2) \\
&\leq \alpha\frac{k_2}{k_1+k_2-k}v(S_1 \cup S'') + (1-\alpha - \alpha\frac{k-k_1}{k_1+k_2-k})v(S_2) \\
&\leq \alpha\frac{k_2}{k_1+k_2-k}v(S_1 \cup S'') + (1-\alpha - \alpha\frac{k-k_1}{k_1+k_2-k})\frac{k_2}{k}v(S') \\
&\leq (\alpha\frac{k_2}{k_1+k_2-k} + (1-\alpha - \alpha\frac{k-k_1}{k_1+k_2-k})\frac{k_2}{k})\max(v(S_1 \cup S''), v(S')) \\
&\leq \gamma \cdot \max(v(S_1 \cup S''), v(S'))
\end{aligned}$$

where the second inequality is due to (1). It remains to bound the value of $\gamma$ (the proof can be found in the appendix):

**Claim 4.4** *Let $\gamma = \max_{k_1 \leq k < k_2} \alpha\frac{k_2}{k_1+k_2-k} + (1-\alpha - \alpha\frac{k-k_1}{k_1+k_2-k})\frac{k_2}{k}$. $\gamma \leq \frac{9}{8}$.*

$\square$

## 4.1 A $(\frac{9}{8} + \epsilon)$-Approximation Submodular Valuations Subject to a Knapsack Constraint

**The Algorithm**

1. For each set $L$, $|L| \leq \frac{1}{\epsilon^2}$ such that $C(L) \leq B$, and each set $L' \subseteq L$, $|L'| = \epsilon \cdot |L|$:

   (a) Let $T = L - L'$.
   
   (b) Define a new valuation $v^{L,L'}(S) = v(S|T)$. Let $a = \min_{j \in L} c_j$. Let $M^{L,L'} = M - T - \{j | c_j > \epsilon \cdot a\}$.
   
   (c) For the set of items $M^{L,L'}$, obtain a strict solution to the LP w.r.t. $v^{L,L'}$ and budget $B^{L,L'} = B - C(T)$: $x_{S_1}^{L,L'} = \alpha^{L,L'}$ and $x_{S_2}^{L,L'} = 1 - \alpha^{L,L'}$.
   
   (d) Find $S' \subseteq S_2$, $C(S') \leq B^{L,L'}$ such that $v^{L,L'}(S') \geq \frac{B^{L,L'}(1-\epsilon)}{C(S_2)} v(S_2)$. If there is no such $S'$ let $A^{L,L'} = \emptyset$ and continue to the next iteration.
   
   (e) Find $S'' \subseteq S_2 - S_1$, $C(S') \leq B^{L,L'} - C(S_1)$, such that $v^{L,L'}(S') \geq \frac{(B^{L,L'} - C(S_1))(1-\epsilon)}{C(S_2 - S_1)} v^{L,L'}(S_2)$. If there is no such $S'$ let $A^{L,L'} = \emptyset$ and continue to the next iteration.
   
   (f) Let $A^{L,L'} = \arg\max(v(T \cup S_1 \cup S''), v(T \cup S'))$.

2. Output $\arg\max_{L,L'} v(A^{L,L'})$.

Notice that if $\epsilon$ is constant then there are only polynomially many iterations. Since each iteration needs polynomially many demand queries, the total number of demand queries is polynomial too. We will use Lemma 4.5 to implement Steps (1d) and (1e), similarly to the algorithm for a cardinality constraint. (Notice that $v^{L,L'}$ is a submodular valuation.)

**Lemma 4.5** *Let $v$ be some submodular valuation, and $S$ be some bundle. Then, there exists a chain $S_1 \subseteq \ldots \subseteq S_{|S|-1} \subseteq S_{|S|} = S$, where $|S_t| = t$ and $v(S_t) \geq \frac{C(S_t)}{C(S)} v(S)$ for every $t$.*

**Proof:** We prove the existence of $S_t$ for $t = |S| - 1$. The lemma will then follow as it implies that we can find $S_1 \subseteq S_2 \subseteq \ldots \subseteq S_{|S|-1} \subseteq S_{|S|} = S$, where $|S_t| = t$, and for each $l$, $|S_l| + 1 = |S_{l+1}|$ and $v(S_l) \geq \frac{C(S_l)}{C(S_{l+1})} v(S_{l+1})$. Now we will have that

$$v(S_t) \geq \frac{C(S_t)}{C(S_{t+1})} v(S_{t+1}) \geq \frac{C(S_t)}{C(S_{t+1})} \cdot \frac{C(S_{t+1})}{C(S_{t+2})} \cdot \ldots \cdot \frac{C(S_{|S|-1})}{C(S_{|S|})} v(S) = \frac{C(S_t)}{C(S)} v(S)$$

Notice that given $v(S_l)$ we can $v(S_{l-1})$ by considering the $l$ subsets of size $l-1$ and taking the one with the highest value.

We now prove the lemma for $t = |S| - 1$. We have that

$$v(S) = \sum_{j=1}^{|S|} v(\{j\}|\{1,\ldots,j-1\})$$

$$\geq \sum_{j=1}^{|S|} v(\{j\}|S - \{j\})$$

where the inequality holds because $v$ is submodular and has decreasing marginal utilities. In particular we have that for some item $j$, $v(\{j|S - \{j\}\}) \leq \frac{c_j}{C(S)}$, i.e., $v(S - \{j\}) \geq \frac{C(S-\{j\})}{C(S)} v(S)$. Let $S' = S - \{j\}$. This completes the proof of the lemma. $\square$

We only analyze one particular iteration, when $L$ is the set of the $\frac{1}{\epsilon^2}$ items with the highest costs in the optimal solution. Furthermore, let $a_1, ..., a_{|L|}$ be some order on the items of $L$ so that for every $i$, $v(a_i|\{1, ...., a_{i-1}\}) \geq v(a_{i+1}|\{1, ...., a_i\})$. Such order can be obtained by starting from the empty set and greedily taking the item from $L$ with the highest marginal contribution that has not been taken yet. Let $L' = \{a_{(1-\epsilon)|L|}, \ldots a_{|L|}\}$. Observe that by submodularity the value of the optimal unrestricted solution in $M^{L,L'} \cup T$ is at least $(1-\epsilon)OPT$, where OPT is the value of the optimal solution. Also notice that for each item $j \in M^{L,L'}$ we have that $c_j \leq \epsilon B^{L,L'}$ since $B^{L,L'} > |L'| \cdot c_j$.

For Step (1d) note that for some $S_t$ we have by Lemma 4.5 that $C(S_t) \leq B^{L,L'}$ and $v^{L,L'}(S') \geq \frac{B^{L,L'}(1-\epsilon)}{C(S_2)} v(S_2)$. Since for each item $j \in M^{L,L'}$ we have that $c_j \leq \epsilon B^{L,L'}$ we immediately get $C(S_t) \geq B^{L,L'}(1-\epsilon)$ and Step (1d) follows. Step (1e) follows by observing that the marginal valuation $v^{L,L'}(\cdot|S_1)$ is submodular too and applying Lemma 4.5 similar to Step (1d).

**Lemma 4.6** $\frac{\gamma}{1-\epsilon} v^{L,L'}(A^{L,L'}) \geq x_{S_1}^{L,L'} v^{L,L'}(S_1) + x_{S_2}^{L,L'} v^{L,L'}(S_2)$.

**Proof:** Recall that $v^{L,L'}(S''|S_1) \geq \frac{(B-C(S_1))(1-\epsilon)}{C(S_2-S_1)} v^{L,L'}(S_2|S_1)$. That is (also using $C(S_2 - S_1) \leq C(S_2)$):

$$v^{L,L'}(S'' \cup S_1) - v^{L,L'}(S_1) \geq \frac{(B-C(S_1))(1-\epsilon)}{C(S_2)}(v^{L,L'}(S_2 \cup S_1) - v^{L,L'}(S_1))$$

Rearranging and using $v^{L,L'}(S_2 \cup S_1) \geq v^{L,L'}(S_2)$ (since $v$ is monotone) we have:

$$v^{L,L'}(S'' \cup S_1) \geq \left(\frac{B-C(S_1)}{C(S_2)} v^{L,L'}(S_2) + \frac{C(S_1)+C(S_2)-B}{C(S_2)} v^{L,L'}(S_1)\right)(1-\epsilon) \quad (2)$$

$$\begin{aligned}
\alpha \cdot v^{L,L'}(S_1) + (1-\alpha) \cdot v^{L,L'}(S_2) &= \alpha \frac{C(S_2)}{C(S_1)+C(S_2)-B}\left(\frac{C(S_1)+C(S_2)-B}{C(S_2)} v^{L,L'}(S_1)\right. \\
&\quad + \left.\frac{B-C(S_1)}{C(S_2)} v^{L,L'}(S_2)\right) + \left(1-\alpha-\alpha\frac{B-C(S_1)}{C(S_1)+C(S_2)-B}\right) v^{L,L'}(S_2) \\
&\leq \alpha \frac{C(S_2)}{C(S_1)+C(S_2)-B} \frac{v^{L,L'}(S_1 \cup S'')}{1-\epsilon} \\
&\quad + \left(1-\alpha-\alpha\frac{B-C(S_1)}{C(S_1)+C(S_2)-B}\right) v^{L,L'}(S_2) \\
&\leq \alpha \frac{C(S_2)}{C(S_1)+C(S_2)-B} \frac{v^{L,L'}(S_1 \cup S'')}{1-\epsilon} \\
&\quad + \left(1-\alpha-\alpha\frac{B-C(S_1)}{C(S_1)+C(S_2)-B}\right) \frac{C(S_2)}{B} \frac{v^{L,L'}(S')}{1-\epsilon} \\
&\leq \left(\alpha \frac{C(S_2)}{C(S_1)+C(S_2)-B} + \left(1-\alpha-\alpha\frac{B-C(S_1)}{C(S_1)+C(S_2)-B}\right)\frac{C(S_2)}{B}\right) \\
&\quad \cdot \frac{\max(v^{L,L'}(S_1 \cup S''), v^{L,L'}(S'))}{1-\epsilon} \\
&\leq \frac{\gamma}{1-\epsilon} \cdot \max(v^{L,L'}(S_1 \cup S''), v^{L,L'}(S'))
\end{aligned}$$

where $\gamma$ showed to be at most $\frac{9}{8}$ in Claim 4.4. □

Recall that we lost at most $\epsilon \cdot OPT$ by discarding items in $L'$. We therefore have that the value of the solution is at least $\frac{8(1-\epsilon)}{9}(1-\epsilon) \cdot OPT$.

# 5 A 2-Approximation for Subadditive Valuations

We show that there exists a 2-approximation algorithm for maximization subject to cardinality constraint. This is the best ratio achievable with a polynomial number of demand queries even if the valuation is XOS, as we show in Section 6. While this is only a slight improvement over the $(2+\epsilon)$ approximation algorithm for the setting of [9], we then show how to extend this algorithm to provide a $(1+\frac{k}{1-\epsilon})$-approximation for maximization subject to $k$-knapsack constraints. In particular this implies that there exists a $(2 + O(\epsilon))$-approximation algorithm for maximization subject to a knapsack constraint ($k = 1$). Previously, the best bound was $O(\log n)$ [9].

Both algorithms provide the same approximation ratio also for non-monotone subadditive valuations, not just monotone ones. In the appendix we show that when the valuation is non-monotone and submodular, then the integrality gap is 2.

**Theorem 5.1** *The following two algorithms exist:*

- *A 2-approximation algorithm for maximizing a (not necessarily monotone) subadditive function subject to a cardinality constraint that uses polynomially many demand queries.*

- *A $(1+\frac{k}{1-\epsilon})$-approximation algorithm for maximizing a (not necessarily monotone) subadditive function subject to $k$-knapsack constraints that uses polynomially many demand queries, for every constant $\epsilon > 0$ and constant $k$.*

We start with the first algorithm, which uses a simple rounding scheme.

**The Algorithm**

1. Obtain a strict fractional solution $x_{S_1} = \alpha$ and $x_{S_2} = 1 - \alpha$.

2. Arbitrarily divide $S_2$ into sets $U_1, \ldots, U_l$ such that for each $i < l$, $|U_i| = k$ and $|U_l| \leq k$.

3. Output $A = \arg\max(v(S_1), v(U_1), \ldots, v(U_l))$.

Notice that the above algorithm can be implemented with a polynomial number of demand queries.

**Lemma 5.2** *Let $A$ be the bundle that the algorithm outputs. Then, $2v(A) \geq x_{S_1}v(S_1) + x_{S_2}v(S_2)$.*

**Proof:**

$$\begin{aligned}
x_{S_1}v(S_1) + x_{S_2}v(S_2) &\leq x_{S_1}v(S_1) + x_{S_2} \cdot l \cdot \max(v(U_1), \ldots, v(U_l)) \\
&\leq (x_{S_1} + x_{S_2} \cdot l) \max(v(S_1), v(U_1), \ldots, v(U_l)) \\
&\leq (x_{S_1} + x_{S_2}(\frac{k_2}{k} + 1))v(A) \\
&\leq (1 + x_{S_2}\frac{|S_2|}{k})v(A) \\
&\leq (1 + 1)v(A) \\
&= 2v(A)
\end{aligned}$$

where the first inequality holds by subadditivity $v(S_2) \leq \sum_{i=1}^{l} v(U_i)$, the third uses $l \leq \lceil \frac{k_2}{k} \rceil$, the fourth inequality uses $x_{S_1} + x_{S_2} \leq 1$, and the last inequality holds since by the LP we have that $x_{S_2}|S_2| \leq k$. □

## 5.1 An Approximation Algorithm for Subadditive Valuations Subject to $k$-Knapsack Constraints

In this section we study maximization subject to $k$ knapsack constraints. In this problem we have a valuation $v$ and $k$ costs for each item $j$: $c_1^i, \ldots, c_k^i$. Let $C_i(S) = \Sigma_{j \in S} c_j^i$. We also have $k$ budgets $B_1, \ldots, B_k$. The goal is to find a maximum-value bundle $S$ such that for every $i$, $C_i(S) \leq B_i$.

We note that $\frac{e}{e+1}$-approximation algorithms exist for maximizing a *montone submodular* valuation subject to $k$ knapsack constraints (the algorithms use only value queries) [5, 18] (note that our algorithms are for the more general subadditive case). Consider a generalization of the LP for a single knapsack constraint:

Maximize: $\sum_S x_S \cdot v(S)$
Subject to:

- For each constraint $i \in [k]$: $\sum_S x_S \cdot C_i(S) \leq B_i$.
- $\sum_S x_S \leq 1$.
- For each bundle $S$: $x_S \geq 0$.

**Proposition 5.3** *The LP can be solved with a polynomial number of demand queries.*

**Proof:** Once again we take its dual and give a separation oracle to the dual to solve the LP.

Minimize: $\sum_{i=1}^k y_i \cdot B_i + z$
Subject to:

- For each bundle $S$: $z + \sum_{i=1}^k y_i \cdot C_i(S) \geq v(S)$.
- For each constraint $i \in [k]$: $y_i \geq 0$.
- $z \geq 0$.

Similar to the case of single knapsack constraint (Proposition 3.1) we can solve this dual LP by ellipsoid method. The separation oracle for the dual is a demand query $\max_T v(T) - \sum_{i=1}^k y_i \cdot C_i(T)$. □

The final algorithm uses enumeration over "big" items.

**Definition 5.4** *An item $j$ is called* big *if $c_j^i \geq \epsilon \cdot B_i$ for some constraint $i \in [k]$. An item is called* small *otherwise.*

Let $\mathcal{B}$ the set of big items and $\mathcal{W}$ be the set of small items.

**The Algorithm**

1. For each set $T \subseteq \mathcal{B}$ such that for each constraint $i \in [k]$: $C_i(T) \leq B_i$:
   (a) Let $M' = T \cup \mathcal{W}$.
   (b) Obtain fractional solution on items from $M'$.
   (c) Divide each bundle $S$ in the fractional solution into sets, $U_1^S, U_2^S, \ldots, U_{l_S}^S$ sets each satisfying budget constraint each $U_i^S$ respects all budget constraints. Additionally $l_S \leq \lceil \sum_{i=1}^k \frac{C_i(S)}{B_i(1-\epsilon)} \rceil$.

(d) Let $A^T = \arg\max(U_j^S)$.

2. Output $\arg\max_{T \subseteq B} v(A^T)$.

Notice that Step (1c) can be implemented for each $S$ as follows: put the subset $T \cap S$ in $U_1^S$ and add items from $S$ without violating the budget constraint of $U_1^S$. Now divide the rest of the items $S$ into bundles $U_1^S, U_2^S, \ldots, U_{l_S}^S$ so that each such bundle respects the budget constraint and for each $i$, $B_i - C_i(U_r^S) \leq \epsilon \cdot B_i$. Since we only have to handle small items, we get that $l_S \leq \lceil \sum_{i=1}^{k} \frac{C_i(S)}{B_i(1-\epsilon)} \rceil$.

As for the number of demand queries we make, notice that in each set $T$ we consider all items are big. Therefore, for this set to respect each of the budget constraints it must be that $|T| \leq k/\epsilon$, which implies that the number of iterations we make is at most $m^{\frac{k}{\epsilon}}$. Since in each iteration we make a polynomial number of demand queries, if $\epsilon$ is constant the total number of queries we make is indeed polynomial.

**Lemma 5.5** *Let $A$ be the bundle that the algorithm outputs. Then, $(1 + \frac{k}{1-\epsilon})v(A) \geq \sum_S x_S v(S)$.*

**Proof:** Consider the set $A^T$ that was obtained in the iteration where $T$ is exactly the set of big items in the optimal solution $O$. The optimal fraction solution with respect to $W \cup T$ has the same value as the unrestricted fractional solution. We can therefore analyze the approximation ratio for this $A^T$.

$$\begin{aligned}
\sum_S x_S v(S) &\leq \sum_S x_S \cdot l_S \cdot \max(v(U_1^S), v(U_2^S), \ldots, v(U_{l_S}^S)) \\
&\leq v(A^T) \sum_S x_S \cdot l_S \\
&\leq v(A^T) \sum_S x_S (\lceil \sum_{i=1}^{k} \frac{C_i(S)}{B_i(1-\epsilon)} \rceil) \\
&\leq v(A^T) \sum_S x_S (\sum_{i=1}^{k} \frac{C_i(S)}{B_i(1-\epsilon)} + 1) \\
&\leq v(A^T)(1 + \sum_S x_S \sum_{i=1}^{k} \frac{C_i(S)}{B_i(1-\epsilon)}) \\
&= v(A^T)(1 + \sum_{i=1}^{k} \frac{1}{B_i(1-\epsilon)} \sum_S x_S C_i(S)) \\
&\leq v(A^T)(1 + \sum_{i=1}^{k} \frac{1}{B_i(1-\epsilon)} B_i) \\
&\leq v(A^T)(1 + \frac{k}{1-\epsilon})
\end{aligned}$$

where the second inequality holds since by subadditivity $v(S) \leq \Sigma_{i=1}^{l_S} v(U_i^S)$, and the second to last inequality holds since $\sum_S x_S C_i(S) \leq B_i$ (by the constraints of the fractional solution). $\square$

# 6 Lower Bounds

## 6.1 A Tight Lower Bound for XOS Valuations

We show that our algorithms from Section 5 are essentially tight: an approximation ratio of $2 - \epsilon$ requires exponentially many demand queries. We note that proving bounds on the power of algorithms with demand queries turned out to be not an easy task, and in particular very different than showing lower bounds on the power of value queries. For example, when we consider value queries there are only finitely many value queries that we have to consider (no more than the number of subsets of $M$), whereas there is an infinite number of demand queries we have to consider. The crux of our proof is to show that the power of any *arbitrary* demand query is limited in some formal sense, hence exponentially many demand queries are needed to distinguish between the case that the optimal value is 2, and between the case it is $1 + \epsilon$.

We note that that our bound holds also for randomized mechanisms, and in fact holds also for a the class of fractionally subadditive valuations, a strict subclass of subadditive valuations.

**Theorem 6.1** *Let $A$ be a randomized algorithm that achieves a $(2 - \epsilon)$-approximation, for some fixed $\epsilon > 0$. Let $\alpha$ be such that $m = k^\alpha$. $A$ makes in expectation at least $\dfrac{\frac{1}{2} \cdot k^{\frac{\epsilon k}{100}(\alpha-1)}}{m(\frac{400e}{\epsilon^2})^{\frac{\epsilon k}{100}} (2e)^{k - \frac{\epsilon k}{100}}}$ demand queries, even if the valuation is XOS.*

In particular, fix some $\epsilon, \gamma > 0$, and let $k = m^{1-\gamma}$. The theorem says that obtaining a $2 - \epsilon$ approximation requires exponential number of demand queries. The following three families of additive valuations will be used in the proof:

1. For every item $j$, define the valuation $I_j$ such that $I_j(\{j\}) = 1$ and for every $t \neq j$, $I_j(\{t\}) = 0$.

2. For every subset $T$, $|T| = k$, , define the valuation $G_T$ such that $G_T(\{j\}) = \frac{2}{k}$ if $j \in T$, and for every $j \notin T$, $G_T(\{j\}) = 0$.

3. The valuation $B$ such that for every item $j$, $B(\{j\}) = \frac{1+\epsilon}{k}$.

For every $T$, $|T| = k$, let $v_T(S) = \max(I_1(S), \ldots, I_m(S), B(S), G_T(S))$. In addition, let $v_\emptyset(S) = \max(I_1(S), \ldots, I_m(S), B(S))$. Notice that the valuations are XOS (i.e., fractionally subadditive) by definition. We will prove that:

**Lemma 6.2** *Determining whether the valuation is $v_\emptyset$ (and not $v_T$, for some $T$) requires in expectation at least $\dfrac{\frac{1}{2} \cdot k^{\frac{\epsilon k}{100}(\alpha-1)}}{m(\frac{400e}{\epsilon^2})^{\frac{\epsilon k}{100}} (2e)^{k - \frac{\epsilon k}{100}}}$ demand queries.*

Lemma 6.2 implies Theorem 6.1: the bundle $T$ is of size $k$ and $v_T(T) = 2$. On the other hand, for every bundle $S$ of size $k$ it holds that $v_\emptyset(S) = 1 + \epsilon$.
**Proof:** Specifically, we show that any deterministic algorithm must make in expectation the specified number of demand queries to determine if the valuation is $v_\emptyset$, when the valuation is chosen uniformly at random from the set $\{v_\emptyset\} \cup_{T:|T|=k} v_T$. Yao's principle delievers the lemma now.

We use the following definition:

**Definition 6.3** *Fix a demand query $dq = (p_1, ..., p_m)$. We say that $dq$ covers $v_T$ if the demand set of $dq$ in $v_T$ contains a set $S$ such that $G_T$ is the maximizing clause of $S$.*

**Claim 6.4** Let $dq = (p_1, ..., p_m)$ be a demand query, and let $v_T$ be a valuation that is covered by $dq$. Let $S \subseteq T$ be some bundle in the demand set of $dq$ in $v_T$ such that $G_T$ is the maximizing clause of $S$. Then, $\Sigma_{j \in S} p_j \leq \frac{2|S|}{k} - 1 + \frac{2}{k}$. Furthermore, we have that $|S| \geq \frac{k}{2} - 1$.

**Proof:** Let $t = \arg\min_{j \in S} p_j$. $G_T$ (and not $I_t$) is the maximizing clause for $S$, and thus:

$$G_T(S) - \Sigma_{j \in S} p_j = \frac{2|S|}{k} - \Sigma_{j \in S} p_j \geq 1 - p_t = I_t(t) - p_t$$

Since $S$ is in the demand set and $G_T$ is its maximizing clause, for each $j \in S$, $p_j \leq \frac{2}{k}$ (otherwise $S - \{j\}$ is more profitable than $S$, and thus $S$ is not in the demand set).

$$\frac{2|S|}{k} - \Sigma_{j \in S} p_j \geq 1 - \frac{2}{k}$$

Reorganizing we have that $\frac{2|S|}{k} - 1 + \frac{2}{k} \geq \Sigma_{j \in S} p_j$, as needed. As for the second part of the claim, notice that if $|S| < \frac{k}{2} - 1$ then $v_T(S) = G_T(S) < 1 - \frac{2}{k}$, and therefore $G_T$ is not the maximizing clause of $S$ since $I_T(t) - p_t < 1 - \frac{2}{k}$. $\square$

The key observation is that if $v_T$ is not covered by $dq$, then the demand set of $dq$ is identical in $v_\emptyset$ and $v_T$. In particular, the only information that executing $dq$ adds is either to determine that the valuation is some specific $v_T$ (in case $v_T$ is covered by $dq$) or to claim that the valuation is not any of the $v_T$'s that are covered by $dq$. However, to determine whether the valuation is $v_\emptyset$, we need to rule out *all* possible values of $T$. The heart of the proof is the following claim:

**Claim 6.5** Fix a demand query $dq = (p_1, ..., p_m)$. The number of bundles that $dq$ covers is at most $\binom{\frac{4k}{\epsilon}}{\frac{\epsilon k}{100}} \cdot \binom{m - \frac{\epsilon k}{100}}{k - \frac{\epsilon k}{100}}$.

**Proof:** We upper bound the number of valuations $V_T$ that $dq$ covers by describing a set $L$ of items of size at most $|L| \leq \frac{4k}{\epsilon}$ with the property that for every $v_T$ that is covered by $dq$, $|L \cap T| > \frac{\epsilon k}{100}$. The number of these bundles is at most:

$$\sum_{i=\frac{\epsilon k}{100}}^{k} \binom{\frac{4k}{\epsilon}}{i} \binom{m - \frac{4k}{\epsilon}}{k - i} \leq \binom{\frac{4k}{\epsilon}}{\frac{\epsilon k}{100}} \cdot \binom{m - \frac{\epsilon k}{100}}{k - \frac{\epsilon k}{100}}$$

We now describe a process that constructs the set $L$. After that, we prove that $L$ is indeed of the specified size.

1. Let $L_0 = \emptyset$, $\mathcal{T} = \{T | v_T \text{ is covered by } dq\}$, and $i = 0$.

2. While $\mathcal{T} \neq \emptyset$:

   (a) Let $i = i + 1$.
   (b) Select some $T_i \in \mathcal{T}$.
   (c) Let $S_{T_i}$ be a set in the demand set of $dq$ in $v_T$ that is maximized in the clause $G_T$.
   (d) Let $L_i = L_{i-1} \cup S_{T_i}$.
   (e) Let $\mathcal{T} = \{T | v_T \text{ is covered by } dq \text{ and } |L_i \cap T| < \frac{\epsilon k}{100}\}$.

3. Let $L = L_i$.

We now show that $|L| \leq \frac{4k}{\epsilon}$. Specifically, we show that if the number of iterations $t$ is more than $\frac{4}{\epsilon}$ then the profit of $L$ is at least 2. This means the demand set does not contain bundles of size at most $k$ at all, since their value is at most 2. Hence $dq$ covers no valuation $v_T$ (i.e., $v(L) - \Sigma_{j \in L} p_j \geq v(T) - \Sigma_{j \in T} p_j$ for every bundle $T$ of size at most $k$). Thus, if $dq$ covers some valuation $v_T$ it must hold that $t \leq \frac{4}{\epsilon}$.

For each iteration $i$, let $S_i = L_i - L_{i-1}$. Observe that by Claim 6.4, $|S_{T_i}| \geq \frac{k}{2} - 1$ and therefore $|S_i| \geq \frac{k}{2} - 1 - \frac{\epsilon k}{100}$. From the same claim we also have that $\Sigma_{j \in S_{T_i}} p_j \leq \frac{2|S_{T_i}|}{k} - 1 + \frac{2}{k}$. Together this implies that $\Sigma_{j \in S_i} p_j \leq \frac{2(|S_i| + \frac{k \cdot \epsilon}{100})}{k} - 1 + \frac{2}{k}$.

We would like to bound the number of iterations $t$ the process goes on. By our discussion above, the process must stop before the profit of $L$ is bigger than 2 (i.e., $v_\emptyset(L) - \Sigma_j p_j > 2$):

$$
\begin{aligned}
v_\emptyset(L) - \Sigma_j p_j &= \frac{(1+\epsilon)|L|}{k} - \Sigma_{j \in L} p_j \\
&= \frac{(1+\epsilon)\Sigma_{i \leq t}|S_i|}{k} - \Sigma_{i \leq t} \Sigma_{j \in S_i} p_j \\
&\geq \frac{(1+\epsilon)\Sigma_{i \leq t}|S_i|}{k} - \Sigma_{i \leq t}(\frac{2|S_i|}{k} - 1 + \frac{2}{k} + \frac{\epsilon}{100}) \\
&= t \cdot (1 - \frac{\epsilon}{100} - \frac{2}{k}) - (1-\epsilon)\Sigma_{i \leq t}\frac{|S_i|}{k} \\
&\geq t \cdot (1 - \frac{\epsilon}{100} - \frac{2}{k}) - (1-\epsilon) \cdot t \\
&= t \cdot (\epsilon - \frac{\epsilon}{100} - \frac{2}{k}) \\
&\geq t \cdot \frac{\epsilon}{2}
\end{aligned}
$$

where the second to last inequality follows since for each $i$, $|S_i| \leq |T_i| \leq k$. Therefore for the profit of $L$ to be at most 2 it must hold that $t \leq \frac{4}{\epsilon}$. Recall that any iteration we are adding to $L$ at most $k$ items. Therefore $|L| \leq t \cdot k \leq \frac{4k}{\epsilon}$. □

Claim 6.5 implies Lemma 6.2: the total number of bundles of size $k$ is $\binom{m}{k}$. To rule out at least $\frac{\binom{m}{k}}{2}$ possible bundles, the number of demand queries $d$ we have to make is at least (we use the bounds $\binom{n}{r} \geq (\frac{n}{r})^r$, $\binom{n}{r} \leq (\frac{ne}{r})^r$ and use $m = k^\alpha$):

$$
\begin{aligned}
\frac{\frac{1}{2} \cdot \binom{k^\alpha}{k}}{\binom{\frac{4k}{\epsilon}}{\frac{\epsilon k}{100}} \cdot \binom{k^\alpha - \frac{\epsilon \cdot k}{100}}{k - \frac{\epsilon k}{100}}} &\geq \frac{\frac{1}{2} \cdot \frac{k^{\alpha k}}{k^k}}{(e\frac{\frac{4k}{\epsilon}}{\frac{\epsilon k}{100}})^{\frac{\epsilon k}{100}} \cdot (e\frac{k^\alpha - \frac{\epsilon \cdot k}{100}}{k - \frac{\epsilon k}{100}})^{k - \frac{\epsilon k}{100}}} \\
&= \frac{\frac{1}{2} \cdot k^{k(\alpha-1)} \cdot (k - \frac{\epsilon k}{100})^{k - \frac{\epsilon k}{100}}}{(\frac{400e}{\epsilon^2})^{\frac{\epsilon k}{100}} (e k^\alpha)^{k - \frac{\epsilon k}{100}}} \\
&\geq \frac{\frac{1}{2} \cdot k^{k(\alpha-1)} \cdot k^{k - \frac{\epsilon k}{100}}}{(\frac{400e}{\epsilon^2})^{\frac{\epsilon k}{100}} (2e k^\alpha)^{k - \frac{\epsilon k}{100}}} \\
&= \frac{\frac{1}{2} \cdot k^{k \cdot \alpha - \frac{\epsilon \cdot k}{100}}}{(\frac{400e}{\epsilon^2})^{\frac{\epsilon k}{100}} (2e)^{k - \frac{\epsilon k}{100}} \cdot k^{\alpha \cdot k - \frac{\alpha \epsilon k}{100}}} \\
&\geq \frac{\frac{1}{2} \cdot k^{\frac{\epsilon k}{100}(\alpha - 1)}}{(\frac{400e}{\epsilon^2})^{\frac{\epsilon k}{100}} (2e)^{k - \frac{\epsilon k}{100}}}
\end{aligned}
$$

Therefore, until we rule out at least half of the bundles, any additional demand query finds $T$ with probability at most $\frac{1}{d}$. To rule out half of the bundles we have to make at least $d$ queries, hence after $\frac{d}{m}$ queries we find $T$ with probability at most $o(1)$. □

## 6.2 Ruling Out an FPTAS for Submodular Maximization Subject to a Knapsack Constraint

We show that there is no FPTAS for the problem of optimizing a submodlar function subject to a knapsack constraint (an FPTAS is a $(1+\epsilon)$-approximation algorithm that the number of demand queries it makes is $poly(m, \frac{1}{\epsilon})$).

**Theorem 6.6** *Let $A$ be a randomized algorithm that achieves an $\frac{\frac{m}{2}-\frac{1}{3}}{\frac{m}{2}-\frac{1}{2}}$-approximation. $A$ makes in expectation at least $\frac{2}{m} \cdot \binom{\frac{m}{2}}{\frac{m}{4}}$ demand queries.*

Notice that the lower bound on the approximation ratio depends polynomially on $m$ and therefore an FPTAS must achieve a better ratio in time $poly(m, \frac{1}{\epsilon})$. This shows that an FPTAS for this problem is impossible. For proof, we consider the following family of instances. Let $A$ be some set of $\frac{m}{2}$ items, and let $B$ be the set that contains the rest of the items. For each item $j \in A$ let $c_j = \frac{2}{m} - \epsilon$, and for each item $j \in B$ let $c_j = \frac{2}{m} + \epsilon$. Let the total budget be 1. For every $T$ of size $\frac{n}{2}$ such that $|T \cap A| = \frac{m}{4}$ define the following valuations:

$$v_T(S) = \begin{cases} |S|, & |S| < \frac{m}{2}; \\ \frac{m}{2}, & |S| > \frac{m}{2}; \\ \frac{m}{2}, & |S| = \frac{m}{2}, C(S) > 1; \\ \frac{m}{2} - \frac{1}{3}, & |S| = T; \\ \frac{m}{2} - \frac{1}{2}, & \text{otherwise.} \end{cases}$$

$$v_\emptyset(S) = \begin{cases} |S|, & |S| < \frac{m}{2}; \\ \frac{m}{2}, & |S| > \frac{m}{2}; \\ \frac{m}{2}, & |S| = \frac{m}{2}, C(S) > 1; \\ \frac{m}{2} - \frac{1}{2}, & \text{otherwise.} \end{cases}$$

**Lemma 6.7** *Determining whether the valuation is $v_\emptyset$ (and not $v_T$, for some $T$) requires in expectation at least $\frac{2}{m} \cdot \binom{\frac{m}{2}}{\frac{m}{4}}$ demand queries.*

Lemma 6.7 implies Theorem 6.6: if the valuation is $v_T$ then $v_T(T) = \frac{m}{2} - \frac{1}{3}$. Otherwise, for every bundle other $S$ of below the budget it holds that $v_\emptyset(S) \leq \frac{m}{2} - \frac{1}{2}$. **Proof:** We use the following definition:

**Definition 6.8** *Fix a demand query $dq = (p_1, ..., p_m)$. We say that $dq$ covers $v_T$ if the demand set of $dq$ in $v_T$ contains $T$.*

**Claim 6.9** *Let $dq = (p_1, ..., p_m)$ be a demand query, and let $v_T, v_{T'}$ be two valuations that are covered by $dq$. Then either $T \cap A = T' \cap A$ or $T \cap B = T' \cap B$.*

**Proof:** Suppose not. Let $a, b \in T$ and $a', b' \in T'$ be such that $a, b \notin T'$, $a', b' \notin T$, $a, a' \in A$, $b, b' \in B$. Define the following bundles:

$$T_{b'} = T - a + b', T'_a = T' - b' + a$$

If the valuation is $v_T$, then $T$ is in the demand set and therefore $v_T(T) - p(T) \geq v_T(T_{b'}) - p(T_{b'})$. Since $C(T_b) > 1$, we have that $\frac{1}{3} \leq p_{b'} - p_a$.

If the valuation is $v_{T'}$, then we have that $v_{T'}(T') - p(T') \geq v_{T'}(T'_a) - p(T'_a)$. Since $C(T'_a) < 1$ we have that $\frac{1}{6} \geq p_{b'} - p_a$. We have reached a contradiction and the claim follows. $\square$

Observe that if $v_T$ is not covered by $dq$, then the demand set of $dq$ is identical in $v_\emptyset$ and $v_T$. In particular, the only information that executing $dq$ adds is either that the valuation is some specific $v_T$ (in case $v_T$ is covered by $dq$) or provide a proof that the valuation is not any of the $v_T$'s that are covered by $dq$. However, to determine whether the valuation is $v_\emptyset$, we need to rule out all possible values of $T$. Suppose that we have $d$ demand queries that fail. There are $\binom{m/2}{m/4} \cdot \binom{m/2}{m/4}$ possible value of $T$ (half of the items in $T$ come from $A$ and the other half from $B$), and by the claim each demand query can rule out at most $\binom{m/2}{m/4}$ (we keep all items in, say, $A$, and choose $\frac{m}{4}$ items from $B$). Therefore, until we rule out at least half of the bundles, any additional demand query finds $T$ with probability at most $\frac{1}{2\binom{m/2}{m/4}}$. To rule out half of the bundles we therefore have to make in expectation at least $\binom{m/2}{m/4}$ queries, hence after $\frac{2}{m} \cdot \binom{m/2}{m/4}$ queries we find $T$ with probability at most $o(1)$. $\square$

### Acknowledgments

We thank Bobby Kleinberg for helpful discussions.

### References

[1] Maria-Florina Balcan, Avrim Blum, and Yishay Mansour. Item pricing for revenue maximization. In EC'08.

[2] Yair Bartal, Rica Gonen, and Noam Nisan. Incentive compatible multi unit combinatorial auctions. In TARK'03.

[3] Liad Blumrosen and Noam Nisan. 2007. Combinatorial Auctions (a survey). In "Algorithmic Game Theory", N. Nisan, T. Roughgarden, E. Tardos and V. Vazirani, editors.

[4] Liad Blumrosen and Noam Nisan. On the computational power of demand queries. *SIAM J. Comput.*, 39(4):1372–1391, 2009.

[5] Chandra Chekuri, Jan Vondrák, and Rico Zenklusen. Submodular function maximization via the multilinear relaxation and contention resolution schemes. In STOC'11.

[6] Shahar Dobzinski. An impossibility result for truthful combinatorial auctions with submodular valuations. In STOC'11.

[7] Shahar Dobzinski, Noam Nisan, and Michael Schapira. Approximation algorithms for combinatorial auctions with complement-free bidders. In STOC'05.

[8] Shahar Dobzinski, Noam Nisan, and Michael Schapira. Truthful randomized mechanisms for combinatorial auctions. In STOC'06.

[9] Shahar Dobzinski, Christos Papadimitriou, and Yaron Singer. Mechanisms for complement-free procurement. In EC'11.

# A Appendix for Section 4

**Proof of Claim 4.4**

The proof is basically an algebraic manipulation:

$$
\begin{aligned}
\gamma &= \max_{k_1 \leq k \leq k_2} \alpha \cdot \frac{k_2}{k_1 + k_2 - k} + ((1-\alpha) - \alpha \cdot \frac{k - k_1}{k_1 + k_2 - k}) \frac{k_2}{k} \\
&= \max_{k_1 \leq k \leq k_2} \frac{k_2 - k}{k_2 - k_1} \cdot \frac{k_2}{k_1 + k_2 - k} + (\frac{k - k_1}{k_2 - k_1} - \frac{k_2 - k}{k_2 - k_1} \cdot \frac{k - k_1}{k_1 + k_2 - k}) \frac{k_2}{k} \\
&= \max_{k_1 \leq k \leq k_2} \frac{k_2 - k}{k_2 - k_1} \cdot \frac{k_2}{k_1 + k_2 - k} \cdot (1 - \frac{k - k_1}{k}) + \frac{k - k_1}{k_2 - k_1} \cdot \frac{k_2}{k} \\
&= \max_{k_1 \leq k \leq k_2} \frac{k_2 - k}{k_2 - k_1} \cdot \frac{k_2}{k_1 + k_2 - k} \cdot \frac{k_1}{k} + \frac{k - k_1}{k_2 - k_1} \cdot \frac{k_2}{k} \\
&= \max_{k_1 \leq k \leq k_2} \frac{k_2}{k(k_2 - k_1)} \cdot \frac{k(k_1 + k_2) - k^2 - k_1^2}{k_1 + k_2 - k} \\
&= \max_{k_1 \leq k \leq k_2} \frac{k_2}{(k_2 - k_1)} \cdot (1 - \frac{k_1^2}{k(k_1 + k_2 - k)}) \\
&= \max_{k_1 \leq k_2} \frac{k_2}{(k_2 - k_1)} \cdot (1 - \frac{4k_1^2}{(k_1 + k_2)^2}) \quad (3) \\
&= \max_{k_1 \leq k_2} \frac{k_2}{(k_2 - k_1)} \cdot \frac{k_2^2 + 2k_1 k_2 - 3k_1^2}{(k_1 + k_2)^2} \\
&= \max_{k_1 \leq k_2} \frac{k_2}{(k_2 - k_1)} \cdot \frac{(k_2 + 3k_1) \cdot (k_2 - k_1)}{(k_1 + k_2)^2} \\
&= \max_{k_1 \leq k_2} \frac{(k_2 + 3k_1) \cdot k_2}{(k_1 + k_2)^2} \\
&= \max_{x \geq 1} \frac{(x + 3) \cdot x}{(x + 1)^2} \\
&= \frac{9}{8}
\end{aligned}
$$

where in (3) we use the fact that the expression is maximized at $k = \frac{k_1 + k_2}{2}$. In the second to last equation we set $x = \frac{k_2}{k_1}$.

# B Integrality Gaps

### An Integrality Gap of $\frac{9}{8}$ for Monotone Submodular Valuations

**Theorem B.1** *There exists a monotone submodular valuation $v$ for which the integrality gap for maximization subject to a cardinality constraint is $\frac{9}{8}$.*

**Proof:** Towards defining the valuation $v$, consider a six-element universe $U = \{e_1, \ldots, e_6\}$. Define the following four sets: $m_1 = \{e_1, e_2, e_3\}$, $m_2 = \{e_1, e_4\}$, $m_3 = \{e_2, e_5\}$, $m_4 = \{e_3, e_5\}$. Now define the following valuation where the set of item is $M = \{m_1, m_2, m_3, m_4\}$. Let $v$ be the following valuation: $v(S) = |\cup_{m_i \in S} m_i|$, i.e., the number of elements in $U$ that the union of the sets in $S$ covers. This valuation is submodular.

Now let the cardinality constraint be $k = 2$. The optimal solution is to take any set $S$ of size 2. The value of the optimal integral solution is 4. On the other hand, consider the following fractional solution: $x_{m_1} = \frac{1}{2}$ and $x_{m_2,m_3,m_4} = \frac{1}{2}$. The fractional solution respects the cardinality constraint and has a value of 4.5. The integrality gap is therefore $\frac{9}{8}$. □

**An Integrality Gap of 2 for Non-Monotone Submodular Valutions**

**Theorem B.2** *For every constant $\epsilon > 0$, there exists a non-monotone submodular valuation $v$ for which the integrality gap for maximization subject to a cardinality constraint is $2 - \epsilon$.*

**Proof:** Let the items be $\{1, \ldots, k^2+1\}$. Define the following non-monotone submodular valuation:

$$v(S) = \begin{cases} 1 - \frac{|S|}{k^2}, & S \subseteq \{2, 3, \ldots, k^2+1\}, 1 \in S; \\ \frac{|S|}{k}, & S \subseteq \{2, 3, \ldots, k^2+1\}. \end{cases}$$

Then the optimal integral solution has a value of 1, whereas the fractional solution $x_{\{1\}} = \frac{k}{k+1}$, $x_{\{2,3,\ldots,k^2+1\}} = \frac{1}{k}$ has value of $1 \cdot \frac{k}{k+1} + k^2 \cdot \frac{1}{k+1} = \frac{2k}{k+1}$. As $k$ tends to infinity the integrality gap tends to 2. □